\documentclass[11pt,a4paper]{article}
\usepackage[margin=2.4cm]{geometry}
\usepackage{amsmath}
\usepackage{graphicx}
\usepackage{booktabs}
\usepackage[font=small,labelfont=bf]{caption}
\usepackage{xcolor}
\usepackage{iftex}
\ifPDFTeX
  \usepackage[T1]{fontenc}
  \usepackage{amssymb}
  \usepackage{newtxtext,newtxmath}      
\else
  \usepackage{fontspec}
  \usepackage{unicode-math}
\fi
\usepackage[hidelinks]{hyperref}
\newcommand{\Sx}{S_{x}}
\newcommand{\Sy}{S_{y}}
\newcommand{\Sz}{S_{z}}

\title{\bfseries V-type Atomic Variance Squeezing}
\author{Chenrui Bi,\ \ Zhao-Yang Tong,$^{\ast}$\ \ Hong-Mei Zou,$^{\dagger}$\ \ Zijin Liang\\[4pt]
\small Synergetic Innovation Center for Quantum Effects and Application, Key Laboratory of\\
\small Low-dimensional Quantum Structures and Quantum Control of Ministry of Education, and\\
\small Hunan Research Center of the Basic Discipline for Quantum Effects and Quantum Technologies,\\
\small School of Physics and Electronics, Hunan Normal University, Changsha 410081, China}
\date{}

\begin{document}
\maketitle

\begin{abstract}
\noindent
Variance squeezing of a V-type three-level atom coupled to a dissipative cavity is studied
analytically and numerically. We first show why the variance criterion is silent in this
system unless the two excited states carry a definite relative phase: for real probability
amplitudes the $|A\rangle$--$|B\rangle$ coherence is real, $\langle \Sz\rangle$ vanishes
identically, and the reference noise level on the right-hand side of the Heisenberg relation
disappears with it. Retaining a phase $\phi$ restores the criterion and reduces the whole
problem to closed form. For an atom prepared in the $\{|A\rangle,|B\rangle\}$ subspace the
$\Sx$ component is squeezed if and only if $\tan\alpha<|\sin\beta\sin\phi|$; the corresponding
squeezing factor is $\xi^{2}_{x}=\tan\alpha/|\sin\beta\sin\phi|$, its $\Sy$ counterpart is
reciprocal, and $\xi^{2}_{x}\xi^{2}_{y}=1/\sin^{2}\!\beta\sin^{2}\!\phi$. The deepest squeezing
according to $V(\Sx)$, obtained for $\alpha=\pi/14$ and $\beta=\phi=\pi/2$, saturates the
Heisenberg relation, transfers the noise
to $\Sy$, and reaches $\xi^{2}_{x}=\tan(\pi/14)=0.228$, i.e.\ $6.4$~dB below the standard
quantum limit. Because the squeezing is present already in the initial state, the dynamics can
only degrade it: population leaking into $|C\rangle$ enters the $\Sx$ variance directly, so
the squeezing factor relaxes to its maximum rather than to zero. Weak atom--cavity coupling
prolongs the lifetime by more than an order of magnitude, and atom--cavity detuning prolongs it
further, by a factor close to $1+\Delta^{2}/\kappa^{2}$; both act by slowing the leakage, and
the detuning route does so by freezing the dynamics rather than by protecting the state. The
SGI parameter is found to redistribute the decay between the symmetric and antisymmetric
$|A\rangle$--$|B\rangle$ combinations without appreciably changing the lifetime, its influence
being proportional to $\gamma_{0}/\kappa$.
\end{abstract}

\section{Introduction}

Quantum noise sets the fundamental limit on the precision with which any observable can be
measured, and squeezing evades that limit by redistributing the fluctuations between two
conjugate observables so that one of them falls below the standard quantum limit (SQL). The
idea was first exploited for the light field, where it suppresses the shot noise of
interferometric phase measurements~\cite{caves}, and was subsequently realized in four-wave
mixing~\cite{slusher} and in parametric down-conversion~\cite{wu}. The same construction
applies to the internal degrees of freedom of an atom, whose spin components form a set of
non-commuting observables: for two Hermitian operators $A$ and $B$ with $[A,B]=iC$, the
Heisenberg relation $(\Delta A)^{2}(\Delta B)^{2}\ge|\langle C\rangle|^{2}/4$ implies that
the atom is squeezed in $A$ whenever $(\Delta A)^{2}<|\langle C\rangle|/2$~\cite{heisenberg}.
Predicted originally for resonance fluorescence~\cite{wallszoller}, atomic squeezing has since
been analysed most thoroughly within the Jaynes--Cummings model, where the atom--field
entanglement generated by the coherent exchange of excitations yields a rich nonclassical
dynamics~\cite{jc1,jc2,scully}. For an ensemble, the collective version of the same
inequality has developed into a metrological resource in its own
right~\cite{wineland1,wineland2,kitagawa,ma}.

Real atoms are rarely ideal two-level systems; they generally possess three or more active
levels, and the additional levels are a resource rather than a complication. Among the
multilevel configurations the V-type atom, in which two excited states share a single ground
state, occupies a special position, because the two excitation and decay amplitudes are not
independent: they interfere. This interference underlies lasing without
inversion~\cite{lwi}, the enhancement of the refractive index in an absorbing
medium~\cite{scullyindex}, spectral-line narrowing and quenching through destructive
interference of the decay channels~\cite{zhuscully}, the spontaneous generation of coherence
between nearly degenerate excited states with non-orthogonal transition
dipoles~\cite{ficekswain}, and electromagnetically induced transparency in the V
configuration~\cite{boller,fleischhauer}. Algebraically, the two dipole operators and the
excited-state coherence operator of a V-type atom form a set of non-commuting observables
embedded in SU(3), so that squeezing can develop in two independent dipole pairs and be
transferred between them in a way that has no two-level analogue.

A realistic cavity, however, is an open system: the field leaks through imperfect mirrors at a
rate $\kappa$, the atom decays from both excited states, and the resulting quantum information
is lost to the environment. These channels set not only the magnitude of the squeezing that
can be reached but also the timescale over which it survives~\cite{wallsbook,gardiner,
carmichael}, and they are the reason why strong-coupling cavity QED is the natural arena for
such studies~\cite{thompson,blais,reiserer}. Dissipation can also be turned into a
resource~\cite{gardiner86,verstraete,cox}, but in the present context it is the adversary.

For the V-type atom specifically, the variance criterion itself is the first difficulty, and it
is here that the existing literature is incomplete. In our recent study of the same
model~\cite{ours}, we found that no variance squeezing whatsoever appears when the atomic
initial state is taken with real probability amplitudes, whereas the $\Sx$ entropy squeezing is
clearly present. The reason is not a physical absence of nonclassicality but a structural
degeneracy of the criterion: real amplitudes make the $|A\rangle$--$|B\rangle$ coherence real,
so that $\langle \Sz\rangle=-2\,\mathrm{Im}\,\rho_{AB}$ vanishes and the SQL $|\langle
\Sz\rangle|/2$ collapses to zero together with the signal it is meant to normalize. The
criterion then compares two vanishing quantities and returns nothing. The standard remedy is
to switch to entropic uncertainty relations~\cite{fang}, which remain informative in this
regime; the alternative explored here is to keep the variance criterion and restore its
reference level. This observation
motivates the present work. By retaining a relative phase between the two excited states we
make $\langle \Sz\rangle$ non-zero, the criterion operational, and the squeezing quantifiable.

Our main results are analytic. For an atom initially in a superposition of $|A\rangle$ and
$|B\rangle$ with populations $\cos^{2}\!\alpha$ and $\sin^{2}\!\alpha\sin^{2}\!\beta$ and a
relative phase $\phi$, we show that $\Sx$ is squeezed at $t=0$ if and only if
$\tan\alpha<|\sin\beta\sin\phi|$, with squeezing factor
$\xi^{2}_{x}=\tan\alpha/|\sin\beta\sin\phi|$. The two components are reciprocally locked,
$\xi^{2}_{x}\xi^{2}_{y}=1/\sin^{2}\!\beta\sin^{2}\!\phi$, so that the familiar ``alternation''
of squeezing between $\Sx$ and $\Sy$ as $\alpha$ is varied is the rotation of a single noise
ellipse rather than two independent effects; the two squeezing intervals touch only for a
minimum-uncertainty state. The optimal configuration, $\alpha=\pi/14$, $\beta=\phi=\pi/2$,
saturates the Heisenberg relation exactly and places the $\Sx$ noise $6.4$~dB below the SQL.
We then follow the decay of this squeezing in the dissipative cavity and identify the
mechanism: the $\Sx$ variance of a V-type atom is, to leading order, the population that does
\emph{not} sit in $|A\rangle$, so the decay of the squeezing is driven by population leaking
into $|C\rangle$ and not only by the loss of coherence. This is why the squeezing factor
relaxes towards unity instead of towards zero.

The paper is organized as follows. Section~\ref{sec:def} fixes the spin operators of the
V-type atom and states the variance criterion, with emphasis on the role of $\langle
\Sz\rangle$ as the reference. Section~\ref{sec:model} recalls the model and the reduced
density matrix. Section~\ref{sec:analytic} derives the closed-form results quoted above.
Section~\ref{sec:results} presents the numerical dynamics and Section~\ref{sec:phys} discusses
the mechanism. Section~\ref{sec:conc} concludes.

\section{Variance criterion for a V-type atom}\label{sec:def}

We label the two excited states $|A\rangle$ and $|B\rangle$ and the ground state $|C\rangle$;
the allowed transitions are $|A\rangle\leftrightarrow|C\rangle$ and
$|B\rangle\leftrightarrow|C\rangle$, while $|A\rangle\leftrightarrow|B\rangle$ is forbidden.
Following Refs.~\cite{riccardi,kurzynski}, the dipole-transition operators read
\begin{equation}
S^{+}=-|A\rangle\langle C|+|C\rangle\langle A|-i|B\rangle\langle C|+i|C\rangle\langle B|,
\qquad S^{-}=(S^{+})^{\dagger},
\end{equation}
and the spin operators are
\begin{equation}
\Sx=\tfrac12(S^{+}+S^{-}),\qquad
\Sy=\tfrac1{2i}(S^{+}-S^{-}),\qquad
\Sz=\tfrac12(S^{+}S^{-}-S^{-}S^{+}).
\end{equation}
In the ordered basis $\{|A\rangle,|B\rangle,|C\rangle\}$ this gives
\begin{equation}
\Sx=\begin{pmatrix}0&0&0\\0&0&-i\\0&i&0\end{pmatrix},\quad
\Sy=\begin{pmatrix}0&0&i\\0&0&0\\-i&0&0\end{pmatrix},\quad
\Sz=\begin{pmatrix}0&-i&0\\i&0&0\\0&0&0\end{pmatrix},
\label{eq:mat}
\end{equation}
which obey $[\Sx,\Sy]=i\Sz$ and cyclic permutations, i.e.\ $[S_{j},S_{k}]=i\varepsilon_{jkl}S_{l}$.

Two features of Eq.~\eqref{eq:mat} deserve emphasis because they control everything that
follows. First, $\Sx$ and $\Sy$ have no matrix element inside the $\{|A\rangle,|B\rangle\}$
subspace: they act as a ``transverse'' pair that connects the excited states to the ground
state. Second, $\Sx^{2}=|B\rangle\langle B|+|C\rangle\langle C|$ and
$\Sy^{2}=|A\rangle\langle A|+|C\rangle\langle C|$, so the variance of either transverse
component counts the population \emph{outside} one of the excited states. The physical content
of this section is a direct consequence of these two facts.

The Heisenberg relation for the pair $(\Sx,\Sy)$ is
\begin{equation}
(\Delta \Sx)^{2}(\Delta \Sy)^{2}\ge\tfrac14|\langle \Sz\rangle|^{2},
\qquad\text{i.e.}\qquad
\Delta \Sx\,\Delta \Sy\ge\tfrac12|\langle \Sz\rangle|,
\label{eq:hup}
\end{equation}
with $\Delta S_{j}=\big(\langle S_{j}^{2}\rangle-\langle S_{j}\rangle^{2}\big)^{1/2}$. The
$S_{j}$ component is variance squeezed when
\begin{equation}
\Delta S_{j}<\Big(\tfrac12|\langle \Sz\rangle|\Big)^{1/2},
\qquad\text{or}\qquad
V(S_{j})\equiv \Delta S_{j}-\Big(\tfrac12|\langle \Sz\rangle|\Big)^{1/2}<0 .
\label{eq:crit}
\end{equation}
Equation~\eqref{eq:crit} is scale free only in its ratio form; for comparing different states
we shall therefore also quote the normalized factor
\begin{equation}
\xi^{2}_{j}\equiv\frac{(\Delta S_{j})^{2}}{\tfrac12|\langle \Sz\rangle|},
\qquad
\text{squeezed}\iff \xi^{2}_{j}<1 ,
\label{eq:xi}
\end{equation}
which is the quantity that can be converted into decibels below the SQL,
$10\log_{10}\xi^{2}_{j}$.

The reference level in Eqs.~\eqref{eq:hup}--\eqref{eq:xi} is $|\langle \Sz\rangle|/2$, and
$\langle \Sz\rangle=-2\,\mathrm{Im}\,\rho_{AB}$ is a purely off-diagonal quantity: it is
non-zero only if the two excited states carry a definite relative phase. This is the formal
origin of the degeneracy mentioned in the introduction. If the amplitudes are real,
$\mathrm{Im}\,\rho_{AB}=0$, the right-hand side of Eq.~\eqref{eq:hup} vanishes, and
Eq.~\eqref{eq:crit} degenerates to ``$\Delta S_{j}<0$'', which is never satisfied but carries no
information either. The same is true of a maximally mixed excited-state manifold, for which the
coherence has decohered. Variance squeezing of a V-type atom is therefore not a property that
can be read off a population distribution: it requires coherence, and it requires that the
coherence be complex.

\section{Model and reduced density matrix}\label{sec:model}

We adopt the model of Refs.~\cite{ours,pan}. A V-type atom sits in a single-mode cavity which
is coupled to a structured environment. Under the rotating-wave
approximation~\cite{agarwal}, and with $\hbar=1$,
\begin{equation}
\begin{split}
\hat H_{0}={}&\omega_{1}|A\rangle\langle A|+\omega_{2}|B\rangle\langle B|
+\omega_{3}|C\rangle\langle C|+\omega_{0}\hat a^{\dagger}\hat a\\
&+\int_{0}^{\infty}\!\eta\,\hat b^{\dagger}(\eta)\hat b(\eta)\,d\eta
+\int_{0}^{\infty}\!\big\{g(\eta)\hat a^{\dagger}\hat b(\eta)+\mathrm{h.c.}\big\}d\eta,
\end{split}
\end{equation}
\begin{equation}
\hat H_{I}=g_{1}\hat a|A\rangle\langle C|+g_{1}^{*}\hat a^{\dagger}|C\rangle\langle A|
+g_{2}\hat a|B\rangle\langle C|+g_{2}^{*}\hat a^{\dagger}|C\rangle\langle B| ,
\end{equation}
where $\hat a$ ($\hat a^{\dagger}$) and $\hat b(\eta)$ ($\hat b^{\dagger}(\eta)$) annihilate
(create) a cavity photon and an excitation of the $\eta$-th environmental mode, respectively,
$g(\eta)=\sqrt{\kappa/\pi}$, $\kappa$ is the spectral width of the environment, and $g_{1}$
($g_{2}$) is the coupling of $|A\rangle$ ($|B\rangle$) to the cavity mode.

The total system starts in
$|\psi\rangle_{0}=\big[a(0)|A\rangle+b(0)|B\rangle+c(0)|C\rangle\big]_{s}\otimes|0\rangle_{e}$
and evolves into
\begin{equation}
|\psi(t)\rangle=\big[a(t)|A\rangle+b(t)|B\rangle+c(t)|C\rangle\big]_{s}\otimes|0\rangle_{e}
+\int d_{\omega}(t)\,|C\rangle_{s}|1_{\omega}\rangle_{e}\,d\omega .
\end{equation}
Assuming a Lorentzian environment centred at $\omega_{0}$, degenerate excited states
($\Delta_{1}=\Delta_{2}\equiv\Delta$, with $\Delta_{i}=\omega_{0}-(\omega_{i}-\omega_{3})$),
and applying the Fano theorem to $i\,\partial_{t}|\psi(t)\rangle=\hat H_{\rm int}(t)|\psi(t)\rangle$
with $\hat H_{\rm int}=e^{i\hat H_{0}t}\hat H_{I}e^{-i\hat H_{0}t}$, the amplitudes are
\begin{equation}
a(t)=\mathcal Q_{1}(t)a(0)+\mathcal Q_{2}(t)b(0),\quad
b(t)=\mathcal Q_{2}(t)a(0)+\mathcal Q_{1}(t)b(0),\quad
c(t)=c(0),
\label{eq:amp}
\end{equation}
with $\mathcal Q_{1,2}=(\mathcal G^{+}\pm\mathcal G^{-})/2$ and
\begin{equation}
\mathcal G^{\pm}(t)=e^{-(\kappa+i\Delta)t/2}
\left[\cosh\!\Big(\frac{R^{\pm}t}{2}\Big)+\frac{\kappa+i\Delta}{R^{\pm}}
\sinh\!\Big(\frac{R^{\pm}t}{2}\Big)\right],
\quad
R^{\pm}=\sqrt{(\kappa+i\Delta)^{2}-2\gamma_{0}(1\pm\theta)\kappa} .
\label{eq:G}
\end{equation}
Here $\gamma_{0}$ is the decay rate of the excited states, and $\theta\in[0,1]$ is the SGI
parameter, fixed by the relative orientation of the two transition dipoles: $\theta=1$ for
parallel dipoles (maximal interference between the two decay channels) and $\theta=0$ for
orthogonal ones. The ratio $\gamma_{0}/\kappa$ separates the Markovian regime
($\kappa\gg2\gamma_{0}$) from the non-Markovian one ($\kappa\ll2\gamma_{0}$)~\cite{bellomo}.
Tracing out the environment gives
\begin{equation}
\rho(t)=\begin{pmatrix}
|a|^{2} & ab^{*} & ac^{*}\\
ba^{*} & |b|^{2} & bc^{*}\\
ca^{*} & cb^{*} & 1-|a|^{2}-|b|^{2}
\end{pmatrix},
\label{eq:rho}
\end{equation}
where $c\equiv c(0)$ and the last diagonal element collects both the $|C\rangle|0\rangle$ and
the $|C\rangle|1_{\omega}\rangle$ contributions; Eq.~\eqref{eq:rho} is therefore the exact
reduction of the pure state and needs no further approximation.

Throughout, $\kappa=1$ sets the unit of time; frequencies are in units of $\kappa$ and times
are given as $\kappa t$.

\section{Closed-form results}\label{sec:analytic}

\subsection{Exact variances and the structure of the criterion}

From Eqs.~\eqref{eq:mat} and \eqref{eq:rho}, the first moments are
\begin{equation}
\langle \Sx\rangle=-2\,\mathrm{Im}\,\rho_{23},\qquad
\langle \Sy\rangle=2\,\mathrm{Im}\,\rho_{13},\qquad
\langle \Sz\rangle=-2\,\mathrm{Im}\,\rho_{12},
\label{eq:mom}
\end{equation}
and, because $\Sx^{2}=1-\rho_{11}$ and $\Sy^{2}=1-\rho_{22}$ in this representation,
\begin{equation}
(\Delta \Sx)^{2}=1-\rho_{11}-4(\mathrm{Im}\,\rho_{23})^{2},
\qquad
(\Delta \Sy)^{2}=1-\rho_{22}-4(\mathrm{Im}\,\rho_{13})^{2}.
\label{eq:var}
\end{equation}
Equation~\eqref{eq:var} is the compact form of the expressions obtained by expanding
$|\rho_{23}|^{2}$ and $\rho_{23}^{2}+\rho_{32}^{2}$; it makes the two contributions
transparent and is what we use below. In the $\{|A\rangle,|B\rangle\}$ subspace, where
$c(0)=0$ and hence $\rho_{13}=\rho_{23}=0$, it simplifies to
\begin{equation}
(\Delta \Sx)^{2}=p_{B}+p_{C},\qquad (\Delta \Sy)^{2}=p_{A}+p_{C},
\label{eq:sub}
\end{equation}
with $p_{i}=\rho_{ii}$. Thus the transverse noise of a V-type atom is the population that
resides outside the corresponding excited state; it is not a coherence at all. Only the
reference level $|\langle \Sz\rangle|/2=|\mathrm{Im}\,\rho_{AB}|$ is of coherent origin. We
stress this point because it is easy to invert: the statement that ``both the variance and the
SQL come from the $|A\rangle$--$|B\rangle$ coherence'' is incorrect for this model, and it
would misplace the mechanism of the decay discussed in Sec.~\ref{sec:phys}.

\subsection{Initial-state dependence}

We take
\begin{equation}
|\psi\rangle_{0}=\big[e^{-i\phi}\cos\alpha\,|A\rangle+\sin\alpha\sin\beta\,|B\rangle
+\sin\alpha\cos\beta\,|C\rangle\big]_{s}\otimes|0\rangle_{e},
\label{eq:init}
\end{equation}
with $\alpha,\beta,\phi\in[-\pi,\pi]$. At $t=0$, Eqs.~\eqref{eq:sub}--\eqref{eq:mom} give
\begin{equation}
(\Delta \Sx)^{2}=\sin^{2}\!\alpha,\qquad
\langle \Sz\rangle=\sin 2\alpha\,\sin\beta\,\sin\phi ,
\label{eq:t0}
\end{equation}
so that $\Sx$ is squeezed, according to Eq.~\eqref{eq:crit}, if and only if
\begin{equation}
\boxed{\ \tan\alpha<|\sin\beta\sin\phi|\ }
\label{eq:main}
\end{equation}
for $\alpha\in(0,\pi/2)$. Three consequences follow at once.

(i) \emph{Only the combination $s\equiv|\sin\beta\sin\phi|$ matters.} The two parameters
$\beta$ and $\phi$ are interchangeable, which is exactly the ``identical influence'' observed
numerically in Figs.~2 and 3; Eq.~\eqref{eq:main} states it analytically.

(ii) \emph{The normalized squeezing factor is closed form.} Dividing
$(\Delta \Sx)^{2}=\sin^{2}\!\alpha$ by $|\langle \Sz\rangle|/2=\sin\alpha\cos\alpha\,s$,
\begin{equation}
\xi^{2}_{x}=\frac{\tan\alpha}{|\sin\beta\sin\phi|},
\qquad
\xi^{2}_{y}=\frac{\cot\alpha}{|\sin\beta\sin\phi|}
\quad(\beta=\pi/2),
\label{eq:xi2}
\end{equation}
where the $\Sy$ expression refers to the $\{|A\rangle,|B\rangle\}$ case $\beta=\pi/2$. In
decibels, $10\log_{10}\xi^{2}_{x}$.

(iii) \emph{The two components are reciprocally locked.}
\begin{equation}
\xi^{2}_{x}\,\xi^{2}_{y}=\frac{1}{\sin^{2}\!\beta\,\sin^{2}\!\phi}\ \ge 1 ,
\label{eq:recip}
\end{equation}
with equality if and only if $|\sin\phi|=1$ and $\beta=\pi/2$. Since
$\xi^{2}_{x}\xi^{2}_{y}$ cannot fall below unity, $\Sx$ and $\Sy$ can never be squeezed
simultaneously, and the ``alternation'' between them as $\alpha$ varies is the rotation of a
single noise ellipse. For $s<1$ the two squeezing intervals do not even touch: from
$\tan\alpha<s$ and $\cot\alpha<s$ the $\Sx$ interval ends at $\alpha=\arctan s$, the $\Sy$
interval starts at $\alpha=\arctan(1/s)$, and the two are separated by a gap
\begin{equation}
\delta\alpha=\frac{\pi}{2}-2\arctan s ,
\label{eq:gap}
\end{equation}
which closes only for a minimum-uncertainty state ($s=1$). This is the precise content of the
``alternating squeezing'' reported previously; for $|\sin\phi|<1$ the alternation is
incomplete.

\subsection{Saturation and the optimal state}

For $\beta=\phi=\pi/2$ the atom is prepared in
$e^{-i\pi/2}\cos\alpha|A\rangle+\sin\alpha|B\rangle$, a pure state of the
$\{|A\rangle,|B\rangle\}$ subspace, and
\begin{equation}
\Delta \Sx=\sin\alpha,\qquad \Delta \Sy=\cos\alpha,\qquad
\tfrac12|\langle \Sz\rangle|=\tfrac12\sin 2\alpha ,
\end{equation}
so that $\Delta \Sx\,\Delta \Sy=\tfrac12\sin2\alpha=\tfrac12|\langle \Sz\rangle|$: the state
saturates Eq.~\eqref{eq:hup} exactly. The squeezing is therefore not a violation of the
uncertainty product but a redistribution within a minimum-uncertainty ellipse, and the
antisqueezing of $\Sy$ is its necessary counterpart. This is the clearest case in which the
V-type atom realizes an ideal squeezed state, and it is why we use it as the reference
configuration.

The depth $V(\Sx)=\sin\alpha-\big(\tfrac12\sin2\alpha\big)^{1/2}$ vanishes at $\alpha=0$ and
$\alpha=\pi/4$ and is minimal in between. Setting $dV/d\alpha=0$ gives
\begin{equation}
4\sin\alpha\cos^{3}\!\alpha=s\cos^{2}\!2\alpha ,
\label{eq:opt}
\end{equation}
whose solution for $s=1$ is $\alpha^{*}=0.2216$, i.e.\ $\alpha^{*}\simeq\pi/14.2$. We adopt
$\alpha=\pi/14$, for which $V(\Sx)=-0.2432$ against $-0.2433$ at the exact optimum, a
difference of $0.05\%$. Table~\ref{tab:opt} collects the numbers.

\begin{table}[t]
\centering
\caption{Variance squeezing of the state
$e^{-i\pi/2}\cos\alpha|A\rangle+\sin\alpha|B\rangle$ ($\beta=\phi=\pi/2$) at $t=0$, with
$\kappa=1$. The SQL column is $\big(\tfrac12|\langle \Sz\rangle|\big)^{1/2}$; the depth is
$10\log_{10}\xi^{2}_{x}$. The last two rows illustrate the cautionary case discussed in the
text: $\xi^{2}_{x}=\tan\alpha$ keeps improving as $\alpha\to0$, but the signal
$\langle \Sz\rangle$ vanishes with it.}
\label{tab:opt}
\begin{tabular}{cccccccc}
\toprule
$\alpha$ & $p_{A}$ & $\langle \Sz\rangle$ & $\Delta \Sx$ & SQL & $V(\Sx)$ & $\xi^{2}_{x}$ & depth (dB)\\
\midrule
$\pi/2$ & 0 & 0 & 1 & 0 & --- & --- & criterion fails\\
$\pi/4$ & 0.500 & 1.000 & 0.707 & 0.707 & 0.000 & 1.000 & 0.0\\
$\pi/14$ & 0.950 & 0.434 & 0.223 & 0.466 & $-0.243$ & 0.228 & $-6.4$\\
$\pi/20$ & 0.976 & 0.309 & 0.156 & 0.393 & $-0.237$ & 0.158 & $-8.0$\\
$\pi/100$ & 0.9995 & 0.063 & 0.031 & 0.177 & $-0.146$ & 0.031 & $-15.0$\\
\bottomrule
\end{tabular}
\end{table}

Two remarks on Table~\ref{tab:opt}. First, maximum polarization is not maximum squeezing: at
$\alpha=\pi/4$ the state is fully polarized, $\langle \Sz\rangle=1$, yet it sits exactly on the
SQL. The optimum is a compromise between keeping $\Delta \Sx=\sin\alpha$ small (small $\alpha$)
and keeping the reference $\tfrac12\sin2\alpha$ large (large $\alpha$), which is why the
optimum lies near $13^{\circ}$ rather than at either end. Second, the two measures in the
table rank the states differently, and the disagreement is instructive. The depth $V(\Sx)$,
which compares the absolute noise with the SQL, has a genuine minimum at $\alpha\simeq\pi/14$.
The normalized factor $\xi^{2}_{x}=\tan\alpha$, by contrast, decreases monotonically as
$\alpha\to0$: formally, arbitrarily many decibels are available at small $\alpha$. But the last
row shows the price: both $\Delta \Sx$ and $\langle \Sz\rangle$ vanish there, so the
``improvement'' of $\xi^{2}_{x}$ is bought with a disappearing signal, exactly the artefact
noted in Ref.~\cite{ours} for the entropy criterion. Neither measure alone is a complete
description; we report $V(\Sx)$ for the dynamics (it is the quantity plotted in the figures)
and quote $\xi^{2}_{x}$ and $|\langle \Sz\rangle|$ alongside it whenever a depth is quoted.

\section{Results}\label{sec:results}

We now follow the evolution of the squeezing in the dissipative cavity, using
Eqs.~\eqref{eq:var}--\eqref{eq:crit} and the amplitudes~\eqref{eq:amp}. Unless stated
otherwise the atom starts in the optimal state and $\theta=0$, $\Delta=0$.

\subsection{Dependence on the initial population parameter \texorpdfstring{$\alpha$}{alpha}}

\begin{figure}[t]
\centering
\includegraphics[width=\textwidth]{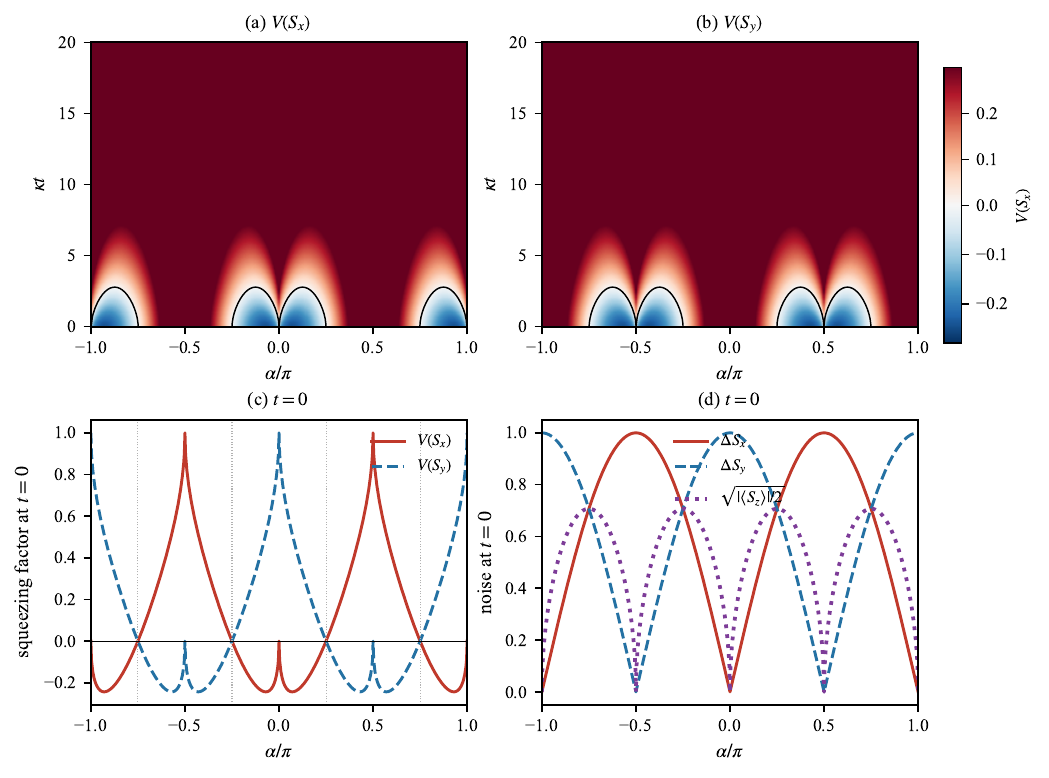}
\caption{Variance squeezing versus the initial-state parameter $\alpha$, for
$\gamma_{0}=0.1$, $\theta=0$, $\Delta=0$, $\beta=\pi/2$ and $\phi=\pi/2$. (a) $V(\Sx)$ and
(b) $V(\Sy)$ as functions of $\kappa t$ and $\alpha$; the colour scale is clipped at zero so
that the squeezed region is white. (c) $V(\Sx)$ (red solid) and $V(\Sy)$ (blue dashed) at $t=0$;
the two curves cross zero at $\alpha=k\pi/4$, so the squeezing alternates with period $\pi$
and each interval has width $\pi/2$. (d) $\Delta \Sx$ (red solid), $\Delta \Sy$ (blue dashed) and the
reference $\big(\tfrac12|\langle \Sz\rangle|\big)^{1/2}$ (purple dotted) at $t=0$; squeezing
corresponds to a curve lying below the reference.}
\label{fig:alpha}
\end{figure}

Figure~\ref{fig:alpha} shows the dynamics for $\beta=\phi=\pi/2$, i.e.\ $s=1$. At $t=0$ the
$\Sx$ component is squeezed for $\alpha\in(k\pi-\pi/4,\,k\pi+\pi/4)$ and the $\Sy$ component
for the complementary intervals, in exact agreement with Eq.~\eqref{eq:main} and
Eq.~\eqref{eq:gap}: the boundaries are $\pm\pi/4$, the intervals have width $\pi/2$, and there
is no gap because the state saturates the uncertainty relation. The deepest $\Sx$ squeezing,
$V(\Sx)=-0.243$, occurs at $\alpha=\pi/14$, and $V(\Sy)=+0.509$ there: the noise removed from $\Sx$
reappears in $\Sy$ in the precise amount required by $\xi^{2}_{x}\xi^{2}_{y}=1$. At
$\alpha=\pi/2$ the atom is in $|B\rangle$, $\langle \Sz\rangle=0$, and the criterion degenerates
again --- the pattern is periodic in $\alpha$ with period $\pi$ and symmetric about
$\alpha=\pi/4$.

For every $\alpha$ the squeezing is maximal at $t=0$ and disappears afterwards; in the
strong-coupling regime the factor shows small non-Markovian revivals, but it never re-enters
the squeezed region. The evolution
never creates squeezing that was not present initially. This is a direct consequence of
Eq.~\eqref{eq:sub}: there is no term in the dynamics that can reduce $p_{B}+p_{C}$ relative to
$|\mathrm{Im}\,\rho_{AB}|$, because both the population transfer to $|C\rangle$ and the decay
of the coherence work in the same direction.

\subsection{Dependence on \texorpdfstring{$\beta$}{beta} and \texorpdfstring{$\phi$}{phi}: the
combination \texorpdfstring{$s=|\sin\beta\sin\phi|$}{s = |sin(beta) sin(phi)|}}

\begin{figure}[t]
\centering
\includegraphics[width=\textwidth]{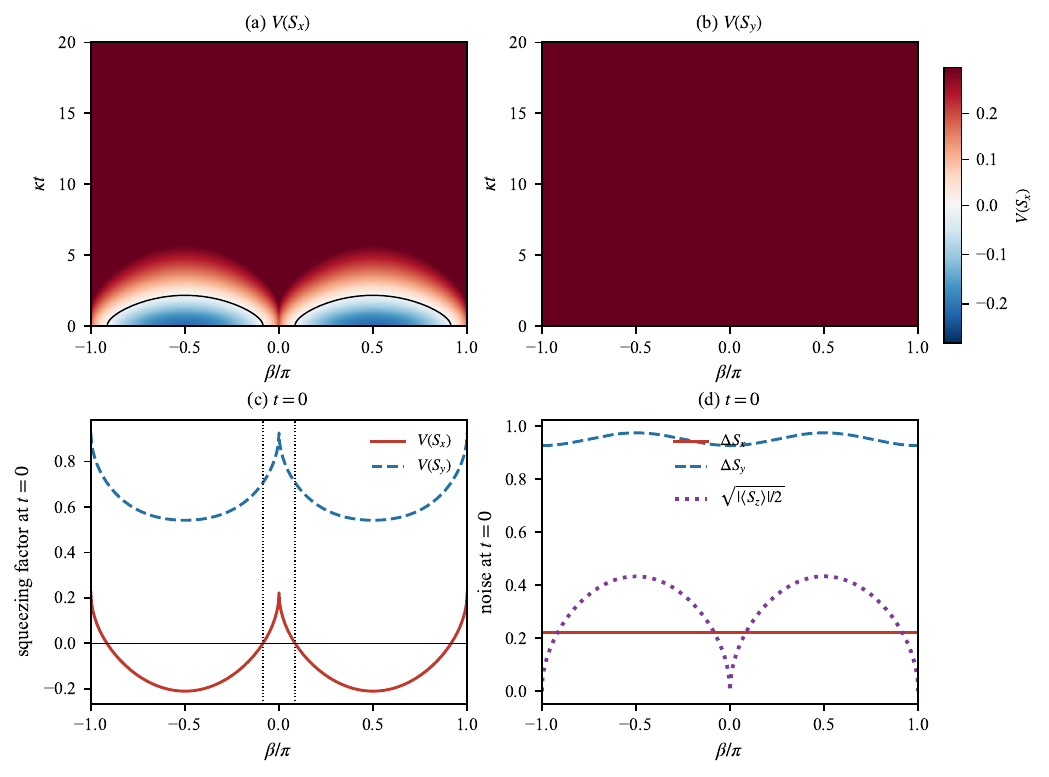}
\caption{As Fig.~\ref{fig:alpha} but versus $\beta$, with $\alpha=\pi/14$, $\phi=\pi/3$,
$\gamma_{0}=0.1$, $\theta=0$ and $\Delta=0$. The boundaries are at
$|\beta|=\arcsin(\tan\alpha/|\sin\phi|)=0.267=\pi/11.8$; the squeezing
interval is $|\beta|\in(0.267,\,2.875)$. Panel (d) shows why the $\Sx$ noise itself (red, flat solid line) does not depend on $\beta$: only the reference level does.}
\label{fig:beta}
\end{figure}

\begin{figure}[t]
\centering
\includegraphics[width=\textwidth]{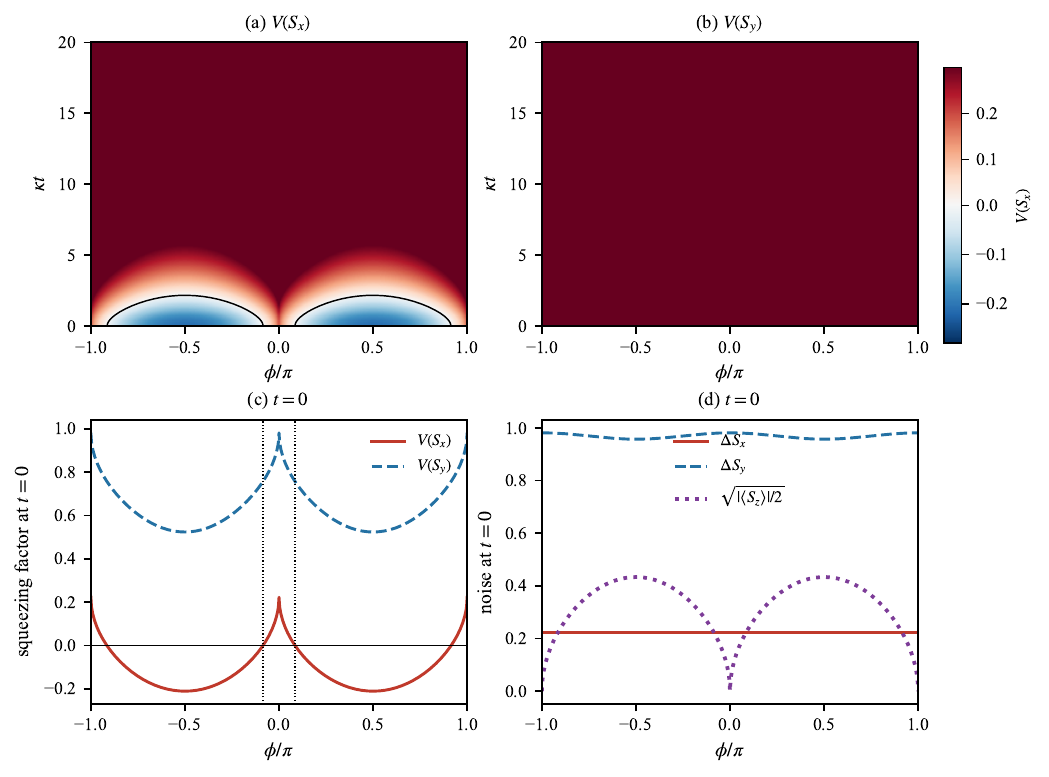}
\caption{As Fig.~\ref{fig:alpha} but versus $\phi$, with $\alpha=\pi/14$, $\beta=\pi/3$,
$\gamma_{0}=0.1$, $\theta=0$ and $\Delta=0$. By Eq.~\eqref{eq:main} the plot is identical to
Fig.~\ref{fig:beta}, since only $|\sin\beta\sin\phi|$ enters; the squeezing interval is
$|\phi|\in(0.267,\,2.875)$.}
\label{fig:phi}
\end{figure}

With $\alpha=\pi/14$ fixed, Eqs.~\eqref{eq:main} and \eqref{eq:xi2} predict squeezing of $\Sx$
for $|\sin\beta|>\tan\alpha/|\sin\phi|=0.264$, i.e.\ $|\beta|>0.267$. This is what
Figs.~\ref{fig:beta} and \ref{fig:phi} show, and it also fixes the optimal choice $\beta=\pi/2$
and $\phi=\pi/2$: both maximize $s$ and hence minimize $\xi^{2}_{x}=\tan\alpha/s$. Because the
two parameters enter only through the product, the two figures coincide panel by panel; this
is the analytic statement behind the numerical observation that ``the influence of $\phi$ is
identical to that of $\beta$''. We emphasize that the boundaries quoted here follow from
Eq.~\eqref{eq:main} and the squeezing
intervals are $|\beta|\in(0.267,\,2.875)$ and $|\phi|\in(0.267,\,2.875)$.

\subsection{Dependence on the atom--cavity coupling ratio}

\begin{figure}[t]
\centering
\includegraphics[width=\textwidth]{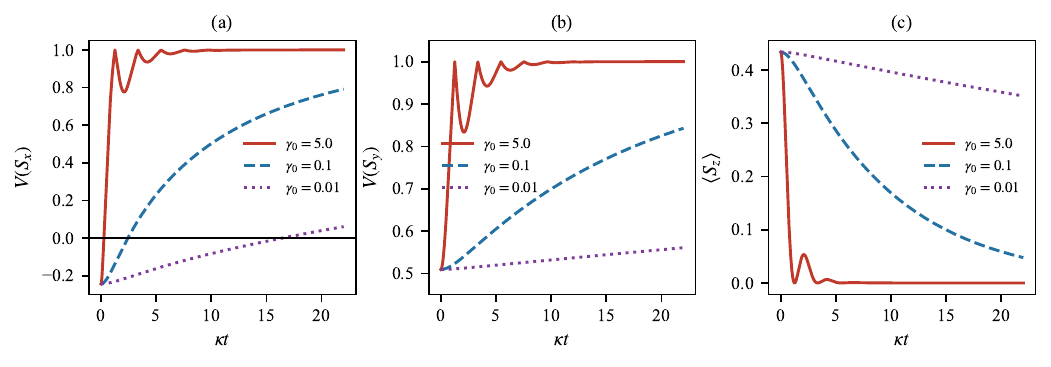}
\caption{Time evolution for $\alpha=\pi/14$, $\beta=\phi=\pi/2$, $\theta=0$, $\Delta=0$ and
$\gamma_{0}=0.01$ (red solid), $0.1$ (blue dashed) and $5$ (purple dotted). (a) $V(\Sx)$;
(b) $V(\Sy)$; (c) $\langle \Sz\rangle$. The initial depth is the same in all three cases; only
the lifetime changes.}
\label{fig:gamma}
\end{figure}

Figure~\ref{fig:gamma} confirms that the initial depth is a property of the state alone: all
three curves start at $V(\Sx)=-0.243$ and $V(\Sy)=+0.509$. What the coupling ratio controls is
the lifetime. The squeezing disappears at $\kappa t=0.26$, $2.43$ and $16.41$ for
$\gamma_{0}=5$, $0.1$ and $0.01$, respectively --- a factor of $64$ between the extremes. The
reason is visible in Fig.~\ref{fig:gamma}(c): $\langle \Sz\rangle$ decays from $0.434$ at a rate
that increases with $\gamma_{0}/\kappa$, and since the criterion is $\Delta \Sx<\big(\tfrac12
|\langle \Sz\rangle|\big)^{1/2}$, the reference falls below the noise as soon as the coherence
decays. We note in passing that the $\Sy$ component is antisqueezed throughout, with
$V(\Sy)\to1$: the noise is conserved, not removed.

\subsection{Dependence on the SGI parameter}

\begin{figure}[t]
\centering
\includegraphics[width=\textwidth]{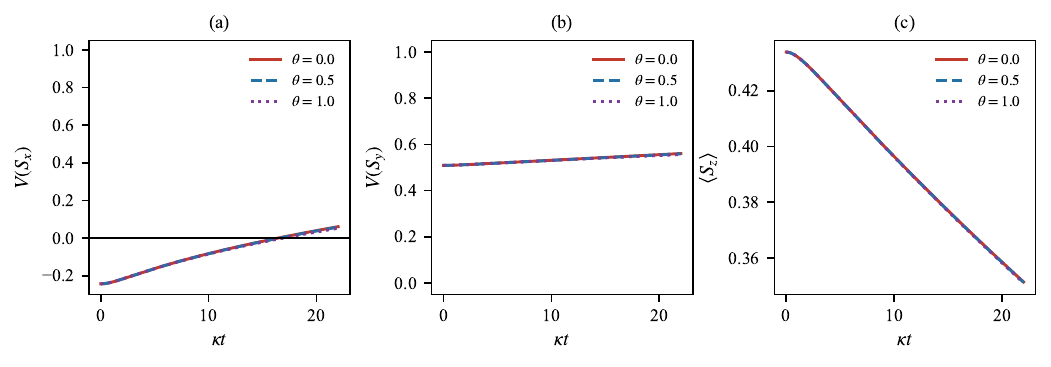}
\caption{Time evolution for $\theta=0$ (red solid), $0.5$ (blue dashed) and $1$ (purple dotted),
with $\alpha=\pi/14$, $\beta=\phi=\pi/2$, $\gamma_{0}=0.01$ and $\Delta=0$. (a) $V(\Sx)$;
(b) $V(\Sy)$; (c) $\langle \Sz\rangle$.}
\label{fig:theta}
\end{figure}

The SGI parameter changes the lifetime by $0.4\%$ only, from $\kappa t=0.257$ at $\theta=1$ to
$0.256$ at $\theta=0$ for $\gamma_{0}=5$, and by a comparable amount in the weak-coupling
regime (Fig.~\ref{fig:theta}). The reason is that $\theta$ enters Eq.~\eqref{eq:G} solely
through $\gamma_{0}(1\pm\theta)\kappa$, i.e.\ it splits the decay rate between the symmetric and
antisymmetric combinations $(a\pm b)/\sqrt2$; its effect on any observable is therefore
proportional to $\gamma_{0}/\kappa$ and is perturbative here. It is worth noting, however, that
this is a statement about the regime and not about the physics: at $\theta=1$ the
antisymmetric combination is a dark state and does not decay at all ($R^{-}=\kappa+i\Delta$,
$\mathcal G^{-}=1$), so that for $\gamma_{0}\gtrsim\kappa$ the residual $|A\rangle$--$|B\rangle$
coherence survives indefinitely and the squeezing freezes at a small but non-zero value instead
of disappearing. This dark-state protection is invisible in Fig.~\ref{fig:theta} only because
$\gamma_{0}=0.01$.

\subsection{Dependence on the atom--cavity detuning}

\begin{figure}[t]
\centering
\includegraphics[width=\textwidth]{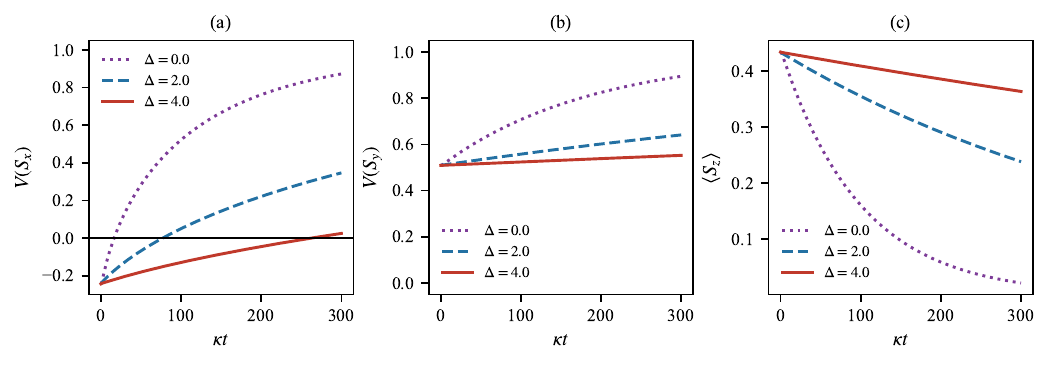}
\caption{Time evolution for $\Delta=0$ (purple dotted), $2$ (blue dashed) and $4$ (red solid),
with $\alpha=\pi/14$, $\beta=\phi=\pi/2$, $\theta=0$ and $\gamma_{0}=0.01$. (a) $V(\Sx)$;
(b) $V(\Sy)$; (c) $\langle \Sz\rangle$.}
\label{fig:det}
\end{figure}

\begin{figure}[t]
\centering
\includegraphics[width=\textwidth]{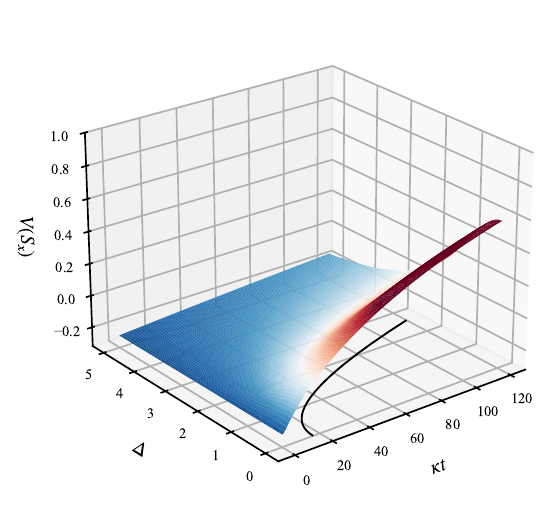}
\caption{$V(\Sx)$ as a function of $\kappa t$ and $\Delta$ for $\alpha=\pi/14$,
$\beta=\phi=\pi/2$, $\theta=0$ and $\gamma_{0}=0.01$. The ridge at $V=0$ moves to longer times
roughly as $1+\Delta^{2}/\kappa^{2}$.}
\label{fig:det3d}
\end{figure}

Detuning is the most effective knob: the squeezing survives until $\kappa t=16.41$, $76.97$ and
$262.49$ for $\Delta=0$, $2$ and $4$ (Figs.~\ref{fig:det} and \ref{fig:det3d}), the lifetime
growing approximately as $1+\Delta^{2}/\kappa^{2}$. Negative detuning acts identically, because
$V(\Sx)$ depends on $\Delta$ only through $|R^{\pm}|$ and is therefore even in $\Delta$.

The mechanism deserves a precise statement, because it is easy to describe it as
``protection''. For $\Delta\gg\kappa$ the arguments of Eq.~\eqref{eq:G} simplify:
$R^{\pm}\simeq\kappa+i\Delta$ and $\mathcal G^{\pm}\to1$, so that $a(t)\simeq a(0)$ and
$b(t)\simeq b(0)$: the atomic state simply stops evolving. The squeezing survives because
nothing happens to it, not because the environment has been made harmless. The effective decay
rate, $\gamma_{\rm eff}\simeq\gamma_{0}\kappa^{2}/(\kappa^{2}+\Delta^{2})$, is suppressed by
the reduced spectral overlap between the atomic transition and the cavity--environment
continuum, and this is the same mechanism that freezes the useful dynamics. A device operated
at $\Delta=4\kappa$ therefore preserves the squeezing for $\kappa t\simeq260$ but also takes
that long to do anything else with it. For applications one should rather work at moderate
detuning, where the lifetime gain is already substantial ($\kappa t\simeq77$ at
$\Delta=2\kappa$) while the dynamics remains fast.

\subsection{Phase diagram of the initial state}

\begin{figure}[t]
\centering
\includegraphics[width=\textwidth]{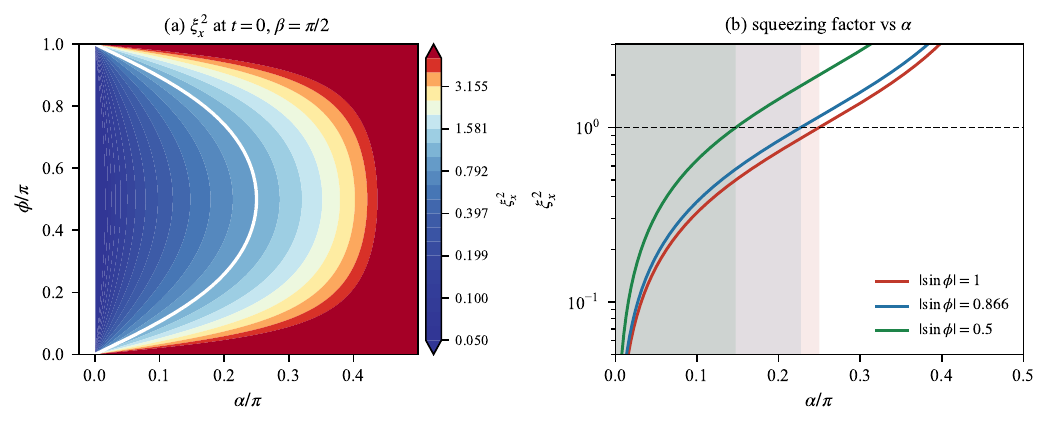}
\caption{(a) $\xi^{2}_{x}$ at $t=0$ in the $(\alpha,\phi)$ plane for $\beta=\pi/2$; the white
curve is the analytic boundary $\tan\alpha=|\sin\phi|$ of Eq.~\eqref{eq:main}, and the dashed
contour marks $\xi^{2}_{x}=1$. (b) $\xi^{2}_{x}$ versus $\alpha$ for
$|\sin\phi|=1$ (red), $0.866$ (blue) and $0.5$ (green); the shaded bands mark the squeezed
regions, whose width shrinks and which no longer touch as $s$ decreases, in agreement with
Eq.~\eqref{eq:gap}.}
\label{fig:phase}
\end{figure}

Figure~\ref{fig:phase} collects the initial-state dependence into a single map. The squeezed
region is bounded by the analytic curve $\tan\alpha=|\sin\phi|$, the depth is largest along
$\phi=\pi/2$, and the alternation between $\Sx$ and $\Sy$ with its gap
$\delta\alpha=\pi/2-2\arctan s$ is visible as the white wedge in panel (b). This replaces the
interval-by-interval description of the earlier literature with a closed-form phase diagram.

\section{Physical discussion}\label{sec:phys}

\emph{Why coherence, and why a phase.} The two quantities entering the criterion
\eqref{eq:crit} have different origins. The noise $(\Delta \Sx)^{2}=p_{B}+p_{C}$ is a
population; the reference $\tfrac12|\langle \Sz\rangle|=|\mathrm{Im}\,\rho_{AB}|$ is a
coherence. Squeezing therefore requires a state that is almost entirely in $|A\rangle$ (so that
$p_{B}+p_{C}$ is small) while still carrying enough $|A\rangle$--$|B\rangle$ coherence to keep
the SQL finite. These two requirements compete, and their competition is what fixes the
optimum at $\alpha\simeq\pi/14$. The phase $\phi$, on the other hand, does not enter
$(\Delta \Sx)^{2}$ at all; it enters only through $\mathrm{Im}\,\rho_{AB}$. This is why a real
initial state cannot show variance squeezing in this system even when it is strongly
nonclassical, and why a phase of $\pi/2$ --- the value that maximizes
$\mathrm{Im}\,\rho_{AB}$ for a given population --- is optimal.

\emph{The ellipse picture.} At $\beta=\phi=\pi/2$ the state saturates Eq.~\eqref{eq:hup}, so
$\xi^{2}_{x}\xi^{2}_{y}=1$ and any reduction of the $\Sx$ noise is paid for exactly by the
$\Sy$ noise. The ``alternating squeezing'' of Figs.~\ref{fig:alpha}--\ref{fig:phase} is the
rotation of this ellipse with $\alpha$: at $\alpha<\pi/4$ the minor axis is along $x$, at
$\alpha>\pi/4$ along $y$. No new physics appears at the crossing; what changes is which
quadrature is short. Away from $|\sin\phi|=1$ the ellipse is no longer minimal, the product
exceeds unity, and a range of $\alpha$ exists in which neither quadrature is squeezed.

\emph{How the squeezing dies.} Equation~\eqref{eq:sub} also explains the shape of the decay.
Population leaking into $|C\rangle$ raises $p_{C}$, hence $\Delta \Sx$, directly --- it does not
have to act through decoherence of the $|A\rangle$--$|B\rangle$ coherence. Since
$\Sx^{2}$ contains $|C\rangle\langle C|$, an atom that has decayed completely into the ground
state has maximal $\Sx$ variance, $\Delta \Sx=1$, and zero reference; the squeezing factor
therefore relaxes to $V=1$, its largest possible value, rather than to zero. The degradation
thus has two contributions of comparable importance, $p_{C}(t)$ growing and
$|\mathrm{Im}\,\rho_{AB}(t)|$ shrinking, and both work in the same direction throughout the
evolution. This is also why no
parameter choice can \emph{create} squeezing after $t=0$: the dynamics has no term that
transfers population back from $|C\rangle$ into a coherent superposition of $|A\rangle$ and
$|B\rangle$.

\emph{Role of $\theta$ and of the environment.} The SGI parameter redistributes the decay
between the $|A\rangle+|B\rangle$ and $|A\rangle-|B\rangle$ channels. For $\theta=1$ the latter
becomes dark and its amplitude is frozen, which is a genuine coherence-protection mechanism,
but its weight in the observable is set by $(\Delta \Sx)^{2}=p_{B}+p_{C}$ and by
$\mathrm{Im}\,\rho_{AB}$, both of which are dominated by the decaying combination when
$\gamma_{0}\ll\kappa$. Hence the weak $\theta$ dependence seen in Fig.~\ref{fig:theta}, and
hence also the prediction that the effect becomes strong once $\gamma_{0}\gtrsim\kappa$.

\emph{Experimental feasibility.} The state of Eq.~\eqref{eq:init} requires a definite relative
phase between the two excited states, i.e.\ a phase-stable superposition of two optical
transitions sharing a common ground state. This is within reach of single-atom cavity QED with
a bichromatic drive acting on the two transitions, and of circuit QED, where a superconducting
qutrit offers the same level structure with the phase controlled directly by the microwave
amplitudes~\cite{blais}. The condition $\gamma_{0}/\kappa\ll1$ corresponds to a bad-cavity
limit in which the atomic coherence is only weakly damped by the cavity, and the required
lifetimes, $\kappa t\sim10$--$10^{2}$, are short compared with typical decoherence times of
such devices. The main practical obstacle is not the lifetime but the phase stability of the
preparation.

\section{Conclusions}\label{sec:conc}

We have re-examined variance squeezing of a V-type atom in a dissipative cavity, with emphasis
on the role of the relative phase between the two excited states. The main conclusions are as
follows.

(i) The variance criterion is structurally degenerate for real probability amplitudes:
$\langle \Sz\rangle=-2\,\mathrm{Im}\,\rho_{AB}$ vanishes, the SQL vanishes with it, and no
squeezing can be detected even when the state is nonclassical. This explains the null result of
Ref.~\cite{ours} and identifies the missing ingredient.

(ii) Retaining a phase $\phi$ makes the problem solvable in closed form. For an atom in the
$\{|A\rangle,|B\rangle\}$ subspace, $\Sx$ is squeezed if and only if
$\tan\alpha<|\sin\beta\sin\phi|$, with $\xi^{2}_{x}=\tan\alpha/|\sin\beta\sin\phi|$ and
$\xi^{2}_{x}\xi^{2}_{y}=1/\sin^{2}\!\beta\sin^{2}\!\phi$. Only the product
$|\sin\beta\sin\phi|$ enters, which is why $\beta$ and $\phi$ have indistinguishable influence.

(iii) The ``alternating'' squeezing of $\Sx$ and $\Sy$ as $\alpha$ is varied is the rotation of
a single noise ellipse. The two squeezing intervals touch only for a minimum-uncertainty state
($|\sin\phi|=1$, $\beta=\pi/2$), where the boundaries are exactly $\alpha=k\pi/4$ and each
interval has width $\pi/2$; otherwise a gap $\pi/2-2\arctan s$ opens between them.

(iv) The deepest squeezing, obtained for $\alpha=\pi/14$ and $\beta=\phi=\pi/2$, saturates the
Heisenberg relation and gives $\xi^{2}_{x}=\tan(\pi/14)=0.228$, i.e.\ $6.4$~dB below the SQL,
with the noise transferred to $\Sy$. Maximum polarization ($\alpha=\pi/4$) gives no squeezing
at all.

(v) The squeezing is present at $t=0$ and can only degrade. Its lifetime is controlled by the
population leaking into $|C\rangle$, which enters $\Delta \Sx$ directly, and by the decay of
$|\mathrm{Im}\,\rho_{AB}|$; the factor $V$ therefore relaxes to unity rather than to zero.

(vi) Weak atom--cavity coupling prolongs the lifetime by a factor of $64$ between
$\gamma_{0}/\kappa=5$ and $0.01$, and detuning prolongs it further, roughly as
$1+\Delta^{2}/\kappa^{2}$; the latter acts by freezing the dynamics rather than by protecting
the state. The SGI parameter has a negligible effect at $\gamma_{0}\ll\kappa$, its influence
scaling with $\gamma_{0}/\kappa$, but it produces dark-state protection at $\theta=1$ and
$\gamma_{0}\gtrsim\kappa$, which is worth a separate study.

These results turn a criterion that is easily degenerate into a quantitative tool: together,
Eqs.~\eqref{eq:main}, \eqref{eq:xi2} and \eqref{eq:recip} give the squeezing of a V-type atom
in terms of three experimentally meaningful parameters, with an explicit optimum and an explicit
lifetime.

\section*{Acknowledgments}
This work was supported by the Foundation Xiangjiang-Laboratory (XJ2302001), Changsha, Hunan,
China, and by the Hunan Research Center of the Basic Discipline for Quantum Effects and Quantum
Technologies, Hunan Normal University, Changsha 410081, China.

\end{document}